# Methods used by WHO to estimate the Global burden of TB disease

Glaziou P*, Sismanidis C*, Pretorius C**, Floyd K*

*Global TB Programme, World Health Organization, Geneva, Switzerland

**Avenir Health, Glastonbury, USA

# Abstract

This paper describes methodological details used by WHO in 2015 to estimate TB incidence, prevalence and mortality. Incidence and mortality are disaggregated by HIV status, age and sex. Methods to derive MDR-TB burden indicators are detailed. Four main methods were used to derive incidence: (*i*) case notification data combined with expert opinion about case detection gaps (120 countries representing 51% of global incidence); (*ii*) results from national TB prevalence surveys (19 countries, 46% of global incidence); (*iii*) notifications in high-income countries adjusted by a standard factor to account for under-reporting and underdiagnosis (73 countries, 3% of global incidence) and (*iv*) capture-recapture modelling (5 countries, 0.5% of global incidence). Prevalence was obtained from results of national prevalence surveys in 21 countries, representing 69% of global prevalence). In other countries, prevalence was estimated from incidence and disease duration. Mortality was obtained from national vital registration systems of mortality surveys in 129 countries (43% of global HIV-negative TB mortality). In other countries, mortality was derived indirectly from incidence and case fatality ratio.

# Introduction

Global targets for reductions in TB disease burden by 2015 were set within the context of the United Nations' Millennium Development Goals (MDGs)[1]. The targets were that TB incidence should be falling, and that TB mortality and prevalence rates should be halved by 2015 compared with their level in 1990. Starting in 1997, global TB reports were published annually by WHO, providing updated data on case notifications and estimated TB burden. In June 2006, the WHO Task Force on TB Impact Measurement was established[2], with the aim of ensuring that WHO's assessment of whether 2015 targets were achieved should be as rigorous, robust and consensus-based as possible. The Task Force reviewed methods and provided recommendations in 2008, 2009 and most recently in March 2015. Methods described in this paper reflect short-term recommendations from the 2015 review. The final MDG assessment was published in WHO's 2015 global TB report[3].

Estimates of the burden of disease caused by TB and measured in terms of incidence, prevalence and mortality are produced annually by WHO using information gathered through surveillance systems (case notifications and death registrations), special studies (including surveys of the prevalence of disease), mortality surveys, surveys of under-reporting of detected TB, in-depth analysis of surveillance and other data, expert opinion and consultation with countries.

# Historical background

Historically, a major source of data to derive incidence estimates were results from tuberculin surveys conducted in children[4]. Early studies showed the following relationship between the annual risk of infection denoted λ and the incidence of smear positive TB denoted $I_{s+}$: one smear positive case infects on average 10 individuals per year for a period of 2 years and a risk of infection of $10^{-2}y^{-1}$ corresponds approximately to an incidence rate of $50 \times 10^{-5}y^{-1}$. However, this relationship no longer holds in the context of modern TB control and in HIV settings[5]. In addition to uncertainty about the relationship between λ and $I_{s+}$, estimates of incidence obtained from tuberculin surveys suffer from other sources of uncertainty and bias, including unpredictable diagnostic performance of the tuberculin test[6], digit preference when reading and recording the size of tuberculin reactions[7], sensitivity to assumptions about reaction sizes attributed to infection[8], sensitivity to the common assumption that the annual risk of infection is age invariant, and lastly, sensitivity of overall TB incidence estimates to the assumed proportion of TB incidence that is smear positive.

A first global and systematic estimation exercise led by WHO in the early 1990s estimated that there were approximately 8 million incident TB cases in 1990 ($152 \times 10^{-5}y^{-1}$) and 2.6-2.9 million deaths ($46\text{-}55 \times 10^{-5}y^{-1}$)[9]. A second major reassessment was published in 1999[10], with an estimated 8 million incident cases for the year 1997 ($136 \times 10^{-5}y^{-1}$), and 1.9 million TB deaths ($32 \times 10^{-5}y^{-1}$). The most important sources of information were case notification data for which gaps in detection and reporting were obtained from expert opinion. In addition, data from 24 tuberculin surveys were translated into incidence and 14 prevalence surveys of TB disease were used.

## Incidence

TB incidence has never been measured through population based surveys at national level because this would require long-term studies among large cohorts of people (hundreds of thousands), involving high costs and challenging logistics. Notifications of TB cases provide a good proxy indication of TB incidence in countries that have both high-performance surveillance systems (for example, there is little under-reporting of diagnosed cases) and where the quality of and access to health care means that few cases remain undiagnosed. In the large number of countries where these criteria are not yet met, better estimates of TB incidence can be obtained from an inventory study. An inventory study is a survey to quantify the level of under-reporting of detected TB cases; if certain conditions are met, capture-recapture methods can also be used to estimate TB incidence [11].

The ultimate goal of TB surveillance is to directly measure TB incidence from national case notifications in all countries. This requires a combination of strengthened surveillance, better quantification of under-reporting (i.e. the number of newly diagnosed cases that are missed by surveillance systems) and universal access to health care (to minimize under-diagnosis of cases). A TB surveillance checklist developed by the WHO Global Task Force on TB Impact Measurement defines the standards that need to be met for notification data to provide a direct measure of TB incidence[12]. By August 2015, a total of 38 countries including 16 high TB burden countries (HBCs) had completed the checklist.

Methods currently used by WHO to estimate TB incidence can be grouped into four major categories. Figure 1 shows the distribution of countries according to the four categories:

1. Case notification data combined with expert opinion about case detection gaps (120 countries);
2. Results from national TB prevalence surveys (19 countries);
3. Notifications in high-income countries adjusted by a standard factor to account for under-reporting and under-diagnosis (73 countries);
4. Capture recapture modelling (5 countries).

## Four main methods

**Method 1 - Case notification data combined with expert opinion about case detection gaps.**

Expert opinion, elicited in regional workshops, national consensus workshops or country missions, is used to estimate levels of under-reporting and under-diagnosis. Trends are estimated using either mortality data, national repeat surveys of the annual risk of infection or exponential interpolation using estimates of case detection gaps for three years. This method was used for 120 countries (Figure 1) that accounted for 51% of the estimated global number of incident cases in 2014. The estimation of case detection gaps is essentially based on an in-depth analysis of surveillance data; experts provide their educated best guess about the range of the plausible detection gap $g$

$$I = \frac{f(N)}{1-g}, g \in [0, 1[$$

where *I* denotes incidence, *N* denotes case notifications, *f* denotes a cubic spline function in countries with large year-to-year fluctuations in *N*, or else, the identity function. The incidence series are completed using assumptions about changes in CFR over time in countries with evidence of improvements in TB prevention and care, such as increased detection coverage over time or improved treatment outcomes, ensuring that the following inequality holds

$$0 \leq \left|\frac{\Delta I}{\Delta t}\right| \leq \left|\frac{\Delta M}{\Delta t}\right|$$

where *M* denotes mortality.

A full description of the methods used in regional workshops where expert opinion was systematically elicited following an in-depth analysis of surveillance data is publicly available in a report of the workshop held for countries in the African Region (in Harare, Zimbabwe, December 2010)[13]. In some countries, case reporting coverage changed significantly during the period 1990-2014 as a result of disease surveillance reforms (e.g. disease surveillance was thoroughly reformed after the SARS epidemic in China, the Ministry of Justice sector notified cases among prisoners in Russia starting in the early 2000s). Trends in incidence were derived from repeat tuberculin survey results in Bhutan, India and Yemen and from trends in mortality in 40 countries (including most countries in Eastern Europe).

The proportion of cases that were not reported were assumed to follow a Beta distribution, with parameters *α* and *β* obtained from the expected value *E* and variance *V* using the method of moments[14], as follows

$$\alpha = E\left(\frac{E(1-E)}{V} - 1\right)$$

$$\beta = (1-E)\left(\frac{E(1-E)}{V} - 1\right)$$

(1)

Time series for the period 1990–2014 were built according to the characteristics of the levels of under-reporting and under-diagnosis that were estimated for the three reference years. A cubic spline extrapolation of *V* and *E*, with knots set at the reference years, was used for countries with low-level or concentrated HIV epidemics. In countries with a generalized HIV epidemic, the trajectory of incidence from 1990 to the first reference year (usually 1997) was based on the annual rate of change in HIV prevalence and time changes in the fraction *F* of incidence attributed to HIV, determined as follows

$$F = \frac{h(\rho - 1)}{h(\rho - 1) + 1} = \frac{\vartheta - h}{1 - h}$$

where $h$ is the prevalence of HIV in the general population, $\rho$ is the TB incidence rate ratio among HIV-positive individuals over HIV-negative individuals and $\vartheta$ is the prevalence of HIV among new TB cases.

If there were insufficient data to determine the factors leading to time-changes in case notifications, incidence was assumed to follow a horizontal trend going through the most recent estimate of incidence.

Limitations of the method based on eliciting expert opinion about gaps in case detection and reporting included a generally small number of interviewed experts; lack of clarity about vested interests when eliciting expert opinion; lack of recognition of over-reporting (due to over-diagnosis, e.g. in some countries of the former Soviet Union implementing a large-scale systematic population screening policy that may result in many people with abnormal chest X-ray but no bacteriological confirmation of TB disease being notified and treated as new TB cases); incomplete data on laboratory quality and high proportion of patients with no bacteriological confirmation of diagnosis are a potential source of error in estimates.

**Method 2 - Results from national TB prevalence surveys.**

Incidence was estimated using prevalence survey results in 19 countries that accounted for 46% of the estimated global number of incident cases in 2014. Two approaches were used to derive incidence from prevalence.

In a first approach, incidence is estimated using measurements from national surveys of the prevalence of TB disease combined with estimates of the duration of disease. Incidence is estimated as the prevalence of TB divided by the average duration of disease assuming epidemic equilibrium: let $N$ denote the size of a closed population with the number of birth and deaths the same for a period $\Delta t>0$, let $C$ be the number of prevalent TB cases, $P$ the prevalence rate so that $P=C/N$. Let $m$ denote the rate of exit from the pool of prevalent cases through mortality, spontaneous self-cure or cure from treatment, and $I$ the rate new cases are added to the pool. At equilibrium during the time period $\Delta t$ and further assuming exponentially distributed durations $d$ such that $d=m^{-1}$

$$I(N - C) = mC$$

$$I = \frac{mC}{N - C} = \frac{P}{d(1 - P)} \approx \frac{P}{d} \qquad (3)$$

In practice, the average duration of presence in the pool of prevalent cases cannot be directly measured. For example, measurements of the duration of symptoms in prevalent TB cases that are detected during a prevalence survey are systematically biased towards lower values, since survey investigations truncate the natural history of undiagnosed disease. Measurements of the duration of disease in notified cases ignore the duration of disease among non-notified cases and are affected by recall biases.

Literature reviews have provided estimates of duration of disease in untreated TB cases from the pre-chemotherapy era (before the 1950s). The best estimate of the mean duration of untreated disease (for smear-positive cases and smear-negative cases combined) in HIV-negative individuals is about three years. There are few data on the duration of disease in HIV-positive individuals. The assumed distributions of disease durations are shown in Table 1.

A second approach consists of estimating disease duration using three model compartments: susceptibles ($S$), untreated TB ($U$) and treated TB ($T$). The size of $U$ and $T$ is obtained from the prevalence survey. Transitions from $U$ to $T$ are determined as follows

$$\frac{dU}{dt} = IS - (\mu_u + \theta_u + \delta)U$$

$$\frac{dT}{dt} = \delta U - (\mu_t + \theta_t)T$$

Where $I$ denotes Incidence, $\mu$ and $\theta$ denote mortality and self-cure or cure (with subscripts $u$ and $t$ indicating untreated and treated cases), respectively, $\delta$ denotes the rate of removal from $U$ through detection and treatment. At equilibrium, the above two equations simplify to

$$I = \frac{U}{d_U}$$

$$\delta U = \frac{T}{d_T}$$

Disease duration (untreated) is obtained from

$$d_U = (1-\pi)\frac{U}{T}d_T$$

where

$$\pi = 1 - \frac{\delta U}{IS}$$

is the proportion of incidence that dies or self-cures before treatment. $\pi$ is assumed distributed uniform with bounds 0 and 0.1. Table 2 shows estimates of incidence from four recent prevalence surveys using this method.

Among limitations of this method is the insufficient power of surveys to estimate the number of prevalent TB cases on treatment with great precision. Further, in most surveys, cases found on treatment during the survey do not have a bacteriological status at onset of treatment documented based on the same criteria as survey cases (particularly when culture is not performed routinely). The method, however, provides more robust estimates of incidence compared with those obtained from expert opinion

(method 1). Figure 2 shows recent changes in estimates as new data became available. As a result of recent survey data in Nigeria and Indonesia, global estimates of incidence increased by about 1 million cases. Time trends, however, were not significantly affected.

In countries with high-level HIV epidemics that completed a prevalence survey, the prevalence of HIV among prevalent TB cases was found systematically lower than the prevalence of HIV among newly notified TB cases, with an HIV prevalence rate ratio among prevalent TB over notified cases ranging from 0.07 in Rwanda (2012) to 0.5 in Malawi (2013). The HIV rate ratio was predicted from a random-effects model fitting data from 5 countries (Malawi, Rwanda, Tanzania, Uganda, Zambia) using a restricted maximum likelihood estimator and setting HIV among notified cases as an effect modifier[15], using the R package metafor[16] (Figure 3). The model was then used to predict HIV prevalence in prevalent cases from HIV prevalence in notified cases in African countries that were not able to measure the prevalence of HIV among survey cases.

The above two methods to derive incidence from prevalence are compared in Table 3. It is not clear which method will perform better. The second method requires a sufficient number of cases on treatment at the time of the survey (as a rule of thumb, at least 30 cases) to generate stable estimates. When both methods can be applied (so far only in low-HIV settings), results from two methods may be combined in a statistical ensemble approach as follows:

The incidence rate obtained using method $i$ is assumed distributed Beta with shape and scale parameters $\alpha_i+1$ and $\beta_i+1$, respectively, and determined using the method of moments based on equation 3: $I_i \sim B(\alpha_i+1,\beta_i+1)$ so that

$$\text{Prob}(x = \text{TB}) = \int_0^1 x B(\alpha_i, \beta_i) \, dx = \frac{\alpha_i + 1}{\alpha_i + \beta_i + 2}$$

The combined probability is then expressed as

$$\text{Prob}(x = \text{TB}) = \frac{\sum \alpha_i + 1}{\sum \alpha_i + \sum \beta_i + 2} \qquad (4)$$

$$\text{Var} = \frac{\left(\sum \alpha + 1\right)\left(\sum \beta + 1\right)}{\left(\sum \alpha + \sum \beta + 2\right)^2 \left(\sum \alpha + \sum \beta + 3\right)}$$

**Method 3 - Notifications in high-income countries adjusted by a standard factor to account for under-reporting and under-diagnosis.**

This method is used for 73 high-income countries (Figure 1), which accounted for 3% of the estimated global number of incident cases in 2014.

TB surveillance systems from countries in the high-income group were assumed to perform similarly well on average. The exceptions were the Republic of Korea, where the under-reporting of TB cases has recently been measured using annual inventory studies and France, where the estimated level of under-reporting was communicated by public health authorities, based on unpublished survey results. In the United Kingdom and the Netherlands, incidence was obtained using capture-recapture modeling (see next section). Surveillance data in this group of countries are usually internally consistent. Consistency checks include detection of rapid fluctuations in the ratio of TB deaths / TB notifications (*M/N* ratio), which may be indicative of reporting problems.

**Method 4 - Capture-recapture modelling.**

This method was used for 5 countries: Egypt[17], Iraq[18], the Netherlands[19], the United Kingdom[20] and Yemen[21]. They accounted for 0.5% of the estimated global number of incident cases in 2014. Capture-recapture modelling was considered in studies with at least 3 lists and estimation of list dependencies[11]. The estimate of the surveillance gap in the UK and the Netherlands was assumed time invariant. In Yemen, trends in incidence were derived from results of two consecutive tuberculin surveys[22]. In Egypt and Iraq, trends were derived using methods described in section describing method 1.

## HIV-positive TB incidence

TB incidence was disaggregated by HIV and CD4 status using the Spectrum software[23]. WHO estimates of TB incidence were used as inputs to the Spectrum HIV model. The model was fitted to WHO estimates of TB incidence, and then used to produce estimates of TB incidence among people living with HIV disaggregated by CD4 category[24]. A regression method was used to estimate the relative risk (RR) for TB incidence according to the CD4 categories used by Spectrum for national HIV projections[25]. Spectrum data used the national projections prepared for the UNAIDS Report on the global AIDS epidemic 2015. The model can also be used to estimate TB mortality among HIV-positive people, the resource requirements associated with recently updated guidance on ART and the impact of ART expansion.

There is no satisfactory way to verify results for TB incidence among people living with HIV when no HIV-testing data in TB are available. A comparative method to disaggregate TB incidence by HIV is shown in annex 2. Provider-initiated testing and counselling with at least 50% HIV testing coverage is the most widely available source of information on the prevalence of HIV in TB patients. However, this source of data is affected by biases, particularly when coverage is closer to 50% than to 100%. As coverage of HIV testing continues to increase globally, biases will decrease.

## Disaggregation by age and sex

Estimates for men (males aged ≥15 years), women (females aged ≥15 years) and children (aged <15 years) are derived as follows. Age and sex disaggregation of smear-positive tuberculosis case notifications has been requested from countries since the establishment of the data collection system in 1995, but with few countries actually reporting these data to WHO. In 2006, the data collection system was revised to additionally monitor age disaggregated notifications for smear-negative and extrapulmonary tuberculosis. The revision also included a further disaggregation of the 0–14 age group category to differentiate the very young (0–4) from the older children (5–14). While reporting of age disaggregated data was limited in the early years of the data collection system, reporting coverage kept improving. For 2012 case notifications, age-specific data reached 99%, 83% and 83% of total smear-positive, smear-negative and extrapulmonary tuberculosis global case notifications. Finally in 2013, another revision of the recording and reporting system was necessary to allow for the capture of cases diagnosed using WHO-approved rapid diagnostic tests (such as Xpert MTB/RIF)[26]. This current revision requests the reporting of all new and relapse case notifications by age and sex. The countries that reported age-disaggregated data in 2014 are shown in Figure 4.

While there are some nationwide surveys that have quantified the amount of under-reporting of cases diagnosed in the health sector outside the network of the NTPs[17,19,27], none have produced precise results by age. Small-scale, convenient-sampled studies indicate that under-reporting of childhood tuberculosis can be very high[28,29] but extrapolation to national and global levels is not yet possible. Plans for implementation of nationwide surveys are under way in selected countries to measure under-reporting of tuberculosis in children[30].

Results from two methods are combined to estimate TB incidence in children, using a statistical ensemble approach based on equation 4. The first method estimates the proportion of all TB cases that are in children as a function of expected age-specific proportions of smear positive TB, according to a previously published approach[31] updated to incorporate recent data[32]. The second method is based on a dynamic model that simulates the course of natural history of TB in children, starting from estimates of tuberculous infection in children as a function of demographic and adult TB prevalence and subsequently modelling progression to pulmonary and extra-pulmonary tuberculosis disease taking into account country-level BCG vaccination coverage and HIV prevalence[33].

Using the sex disaggregated reporting of TB case notification data we calculated the ratio of the number of TB cases notified in men compared with women as a measure of the ratio $r_1$ for incident cases, assuming no sex differential in the detection of incident cases. Evidence from prevalence surveys consistently show bigger recording and detection gaps in men[34]. The assumption of no sex differential in the detection of incident cases may thus lead to underestimating the proportion of men among incident cases. With currently available data, it is not possible to estimate male and female case detection ratios for all countries.

Overall incidence in adults 15 years or over ($I_a$) can be disaggregated into estimates among men ($I_m$) and women ($I_w$) as shown

$$I_m = I_a \frac{r_1}{1+r_1}$$

$$I_w = \frac{I_a}{1+r_1}$$
(6)

where $r_1 = I_m/I_w$ and $I_a = I_m + I_w$.

Producing estimates of TB incidence among children is challenging primarily due to the lack of well performing diagnostics to confirm childhood TB and the lack of age-specific, nationwide, robust survey and surveillance data.

## Prevalence

### Population-based surveys

The best way to measure the prevalence of TB is through national population-based surveys of TB disease[35,36]. Data from such surveys are available for an increasing number of countries and were used for 21 countries (Figure 4), representing 69% of global prevalence in 2014. Measurements of prevalence are typically confined to the adult population, exclude extrapulmonary cases and do not allow the diagnosis of cases of culture-negative pulmonary TB.

TB prevalence all forms and all ages ($P$) is measured as: bacteriologically-confirmed pulmonary TB prevalence ($P_p$) among those aged ≥15 measured from national survey ($P_a$), adjusted for pulmonary TB in children ($P_c$) and the proportion $e$ of extra-pulmonary TB all ages

$$P_p = cP_c + (1-c)P_a$$

where *c* is the proportion of children among the total country population.

$$P = \frac{P_p}{1-e}$$

The estimate of overall prevalence *P* is affected by sampling uncertainty (relative precision is typically about 20%), and uncertainty about *e* (of note, values for *e* vary widely among countries with high-performance TB surveillance) and $P_c$. The quality of routine surveillance data to inform levels of pulmonary TB in children and extra-pulmonary TB for all ages is often questionable.

## Indirect estimates

Indirect estimates of prevalence were calculated by solving equation 3 for *P*, summing over 4 case categories

$$P = \sum I_{i,j} d_{i,j}, i \in \{1,2\}, j \in \{1,2\}$$

where the index variable *i* denotes HIV+ and HIV−, the index variable *j* denotes treated and non-treated cases, and *d* denotes the duration of disease.

When there is no direct measurement from a national survey of the prevalence of TB disease, prevalence is the most uncertain of the three TB indicators used to measure disease burden. This is because prevalence is the sum of products of two uncertain quantities, incidence and disease duration. There is scarce empirical data on disease duration (a typically large proportion of bacteriologically confirmed cases detected during TB prevalence surveys did not report symptoms suggestive of TB at the time of survey investigations[34]). Unless measurements were available from national programmes (for example, Turkey), assumptions of the duration of disease were used as shown in Table 1. An important limitation is that duration is considered time invariant within case categories.

## Mortality

The best sources of data about deaths from TB (excluding TB deaths among HIV-positive people) are vital registration (VR) systems in which causes of death are coded according to ICD-10 (although the older ICD-9 and ICD-8 classification are still in use in several countries), using ICD-10: A15-A19 codes, equivalent to ICD-9: 010-018. When people with AIDS die from TB, HIV is registered as the underlying cause of death and TB is recorded as a contributory cause. Since one third of countries with VR systems report to WHO only the underlying causes of death and not contributory causes, VR data usually cannot be used to estimate the number of TB deaths in HIV-positive people. Two methods were used to estimate TB mortality among HIV-negative people:

- direct measurements of mortality from VR systems or mortality surveys (129 countries, in green in Figure 6);
- indirect estimates derived from multiplying estimates of TB incidence by estimates of the CFR (88 countries).

**Estimating TB mortality among HIV-negative people from vital registration data and mortality surveys**

As of July 2015, 130 countries had reported mortality data to WHO (including data from sample VR systems and mortality surveys), including 10 of the 22 high TB burden countries. The VR data on TB deaths from Zimbabwe were not used because large numbers of HIV deaths were miscoded as TB deaths. Improved empirical adjustment procedures for such miscoding have recently been published[37]. Estimates for South Africa adjusted for HIV/TB miscoding were obtained from the Institute of Health Metrics and Evaluation at http://vizhub.healthdata.org/cod/. Results from mortality surveys were used to estimate TB mortality in India and Viet Nam as an interim measure.

Among the countries for which VR data could be used (Figure 5), there were 2361 country-year data points 1990–2014, after 13 outlier data points from systems with very low coverage (<20%) as estimated by WHO[38] or very high proportion of ill-defined causes (>50%) were excluded for analytical purposes. The median number of data points per country was 21 (Interquartile range 15 - 23).

Reports of TB mortality were adjusted upwards to account for incomplete coverage (estimated deaths with no cause documented) and ill-defined causes of death (ICD-9: B46, ICD-10: R00–R99)[38]. It was assumed that the proportion of TB deaths among deaths not recorded by the VR system was the same as the proportion of TB deaths in VR-recorded deaths. For VR-recorded deaths with ill-defined causes, it was assumed that the proportion of deaths attributable to TB was the same as the observed proportion in recorded deaths. The adjusted number of TB deaths $\kappa_a$ was obtained from the VR report $\kappa$ as follows:

$$\kappa_a = \frac{\kappa}{v(1-g)}$$

where $v$ denotes coverage (i.e. the number of deaths with a documented cause divided by the total number of estimated deaths) and $g$ denotes the proportion of ill-defined causes. The uncertainty related to the adjustment was estimated as follows:

$$\hat{\sigma} = \frac{\kappa}{4}\left[\frac{1}{v(1-g)-1}\right]$$

The uncertainty calculation does not account for miscoding, such as HIV deaths miscoded as deaths due to TB, except in South Africa.

Missing data between existing adjusted data points were interpolated. Trailing missing values were predicted using exponential smoothing models for time series[39]. A penalized likelihood method based on the in-sample fit was used for country-specific model selection. Leading missing values were similarly predicted backwards to 1990.

In 2014, 43% of global TB mortality (excluding HIV) was directly measured from VR or survey data (or imputed from survey or VR data from previous years). The remaining 57% was estimated using the indirect methods described in the next section.

## Estimating TB mortality among HIV-negative people from estimates of case fatality rates and TB incidence

In 88 countries lacking VR data of the necessary coverage and quality, TB mortality was estimated as the product of TB incidence and the case fatality rate (CFR) after disaggregation by case type as shown in Table 4, following a literature review of CFRs by the TB Modelling and Analysis Consortium (TB-MAC):

$$M^- = (I^- - T^-)f_u^- + T^- f_t^- \qquad (5)$$

where $M$ denotes mortality, $I$ incidence. $f_u$ and $f_t$ denote CFRs untreated and treated, respectively and the superscript denotes HIV status. $T$ denotes the number of treated TB cases. In countries where the number of treated patients that are not notified (under-reporting) is known from an inventory study, the number of notified cases is adjusted upwards to estimate $T$ accounting for under-reporting.

Figure 7 shows a comparison of 129 VR-based mortality estimates for 2014 and indirect estimates obtained from the CFR approach for the same countries. Of note, countries with VR data tend to be of a higher socio-economic status compared with countries with no VR data where the indirect approach was used.

## Estimating TB mortality among HIV-positive people

TB mortality among HIV-positive is calculated using equation 5, exchanging superscripts - with +. The case fatality ratios were obtained in collaboration with the TB Modeling and Analysis Consortium (TB-MAC), and are shown in Table 5. The disaggregation of incident TB into treated and not treated cases is based on the ratio of the point estimates for incident and notified cases, adjusted for under-reporting. A single CFR was used for all bootstrapped mortality estimates[24].

Direct measurements of HIV-associated TB mortality are urgently needed. This is especially the case for countries such as South Africa and Zimbabwe, where national VR systems are already in place. In other countries, more efforts are required to initiate the implementation of sample VR systems as an interim measure.

## Disaggregation of TB mortality by age and sex

From the age-specific adjusted (for coverage and ill-defined causes) number of deaths from VR, we first estimate the ratio $r_2$ of rates in children ($M_{0-14}$) compared to adults ($M_{15+}$) (equation 7). The estimation of $r_2$ is based a chained equations multiple imputation approach. The imputation model covariates include total notifications, population proportion aged more than 65 years, an indicator variable for the epidemiological region and whether a country was one of the 22 HBCs[3]. The overall mortality rate for all ages ($M$) can be expressed as a weighted average of mortality in children and adults, where $c$ is the proportion of children among the general population (equation 8)

$$r_2 = \frac{M_{0-14}}{M_{15+}} \tag{7}$$

$$M = cM_{0-14} + (1-c)M_{15+} \tag{8}$$

In countries with VR or mortality survey data, $M_{0-14}$ is directly measured. For the sex disaggregation of TB mortality among adults ($M_{15+}$), sex-specific adjusted (for coverage and ill-defined causes) number of deaths from VR to estimate mortality rates in men $M_m$ and women $M_w$ were used. The ratio of these rates $r_3 = M_m/M_w$ is either directly measured in countries with VR data or imputed in countries without and sex-specific mortality rates are then derived in a manner similar to shown in equation 6.

TB deaths among HIV-positive people were disaggregated by age and sex using the assumption that the child to adult and men to women ratios are the same as the corresponding ratios of AIDS deaths estimated by UNAIDS.

## Projections up to 2015

Projections of incidence, prevalence and mortality up to 2015 enable assessment of whether global targets set for 2015 are likely to be achieved at global, regional and country levels. Projections for the year 2015 were made using exponential smoothing models fitted to data from 2007–2014, based on an algorithm that selects the best among models within a family of exponential smoothing models, using a penalized likelihood method as a selection criterion[40]. Point forecasts are computed using the best model with optimized parameters and uncertainty is propagated using analytical methods described in the next section.

## Estimation of uncertainty

There are many potential sources of uncertainty associated with estimates of TB incidence, prevalence and mortality, as well as estimates of the burden of HIV-associated TB and MDR-TB. These include uncertainties in input data, in parameter values, in extrapolations used to impute missing data, and in the models used. Uncertainty in population estimates was not accounted for.

Notification data are of uneven quality. Cases may be under-reported (for example, missing quarterly reports from remote administrative areas are not uncommon), misclassified (in particular, misclassification of recurrent cases in the category of new cases is common), or over-reported as a result of duplicated entries in TB information systems. The latter two issues can only be addressed efficiently in countries with case-based nationwide TB databases that include patient identifiers. Sudden changes in notifications over time are often the result of errors or inconsistencies in reporting.

Uncertainty bounds and ranges were defined as the 2.5th and 97.5th centiles of outcome distributions. The general approach to uncertainty analyses was to draw values from specified distributions in Monte Carlo simulations or else, uncertainty was propagated analytically by approximating the mean and the second central moment of functions of random variables using higher-order Taylor series expansion[41] in a matrix based approach[42].

# Conclusion

The measurement methods described here can be combined to assess tuberculosis incidence, prevalence and mortality, to evaluate progress towards targets for tuberculosis control and the level of achievement of the MDGs for TB. Alternative TB burden estimation methods have been developed by the Institute of Health Metrics and Evaluation[43], with generally consistent results at the global level compared with WHO, but with marked differences in specific countries. Discrepancies in estimates from different agencies reflect the questionable quality and completeness of the underlying data. Further convergence in estimates will result from improvements in measurements at country level. National control programmes should be able to measure the level and time trends in incidence through well-performing TB surveillance with universal access to health. In countries with incomplete routine surveillance, prevalence surveys of TB disease provide estimates of TB burden that do not heavily rely on expert opinion. The performance of TB surveillance should be assessed periodically[12] and the level of under-reporting should be measured[11] and minimized. Tuberculosis mortality will ideally be measured by counting deaths in a comprehensive vital registration system[38].

The challenge presented by assessing MDGs achievements has been to measure trends in TB incidence, prevalence, and deaths. The MDGs have now been followed by a set of 17 Sustainable Development Goals (SDGs) with an end date of 2030. A target within the health-related SDG is to "End the epidemics of AIDS, TB, malaria and neglected tropical diseases, and combat hepatitis, water-borne diseases and other communicable diseases". WHO's post-2015 global TB strategy, known as the End TB Strategy[44], also has the goal of ending the global TB epidemic, with corresponding targets of a 90% reduction in TB deaths and an 80% reduction in the TB incidence rate by 2030. Improved measurements through substantial investments in health information systems, TB surveillance and the broader SDG agenda will provide a firmer basis for monitoring progress towards the End TB Strategy targets and ultimate TB elimination.

## Acknowledgements
Ibrahim Abubakar, Sandra Alba, Elisabeth Allen, Martien Borgdorff, Jaap Broekmans, Ken Castro, Frank Cobelens, Ted Cohen, Charlotte Colvin, Sarah Cook-Scalise, Liz Corbett, Simon Cousens, Pete Dodd, Katherine Fielding, Peter Godfrey-Faussett, Rein Houben, Li Liu, Mary Mahy, Valérie Schwoebel, Cherise Scott, James Seddon, Andrew Thomson, Edine Tiemersma, Hazim Timimi, Theo Vos, Emilia Vynnycky and Richard White reviewed the described methods to derive TB incidence, prevalence and mortality with disaggregation by age and sex and provided specific recommendations to improve them.

# Annex 1 - Definitions

**Incidence** is defined as the number of new and recurrent (relapse) episodes of TB (all forms) occurring in a given year. Recurrent episodes are defined as a new episode of TB in people who have had TB in the past and for whom there was bacteriological confirmation of cure and/or documentation that treatment was completed.

**Prevalence** is defined as the number of TB cases (all forms) at a the middle of the year.

**Mortality** from TB is defined as the number of deaths caused by TB in HIV-negative people occurring in a given year, according to the latest revision of the International classification of diseases (ICD-10). TB deaths among HIV-positive people are classified as HIV deaths in ICD-10. For this reason, estimates of deaths from TB in HIV-positive people are presented separately from those in HIV-negative people.

The **case fatality rate** is the risk of death from TB among people with active TB disease.

The **case notification** rate refers to new and recurrent episodes of TB notified for a given year. Patients reported in the *unknown history* category are considered incident TB episodes (new or recurrent).

**Population estimates** were the 2015 revision of the World Population Prospects, which is produced by the United Nations Population Division (UNPD, http://esa.un.org/unpd/wpp/). The UNPD estimates sometimes differ from those made by countries.

# Annex 2 - Relationship between HIV prevalence in new TB cases and HIV prevalence in the general population

Let *I* and *N* denote incident cases and the total population, respectively, superscripts + and - denote HIV status, $\vartheta$ is the prevalence of HIV among new TB cases, $h$ is the prevalence of HIV in the general population and $\rho$ is the incidence rate ratio (HIV-positive over HIV-negative).

$$\rho = \frac{I^+/N^+}{I^-/N^-} > 1$$

$$\rho \frac{I^-}{I^+} = \frac{N^-}{N^+}$$

$$\rho \frac{I - I^+}{I^+} = \frac{N - N^+}{N^+}$$

$$\frac{I^+}{I} = \frac{\rho \frac{N^+}{N}}{1 + (\rho - 1)\frac{N^+}{N}} = \vartheta$$

$$\vartheta = \frac{h\rho}{1 + h(\rho - 1)}$$

The TB incidence rate ratio $\rho$ can be estimated by fitting the following linear model with a slope constrained to 1

$$\log(\hat{\rho}) = \log\left(\frac{\vartheta}{1-\vartheta}\right) - \log\left(\frac{h}{1-h}\right), (\vartheta, h) \in \,]0,1[$$

# Annex 3 - Implementation steps

The methods described in the paper were implemented in the following steps:

1. Estimating overall TB incidence after review and cleaning of case notification data;
2. cleaning and adjusting raw mortality data from VR systems and mortality surveys, followed by imputation of missing values in countries with VR or survey data – in some countries, step 1 was updated to account for mortality data;
3. cleaning of measurements of HIV prevalence among TB patients followed by estimating HIV-positive TB incidence using the Spectrum programme and HIV-positive TB mortality;
4. estimating HIV-negative TB mortality in countries with no VR data followed with an update of step 1 in some countries;
5. reviewing prevalence measurements, adjusting for childhood TB and bacteriologically unconfirmed TB, and estimating prevalence followed with an update of step 1 in some countries;
6. estimating incidence and mortality disaggregated by age and sex and disaggregated by drug resistance status.

# Tables

**Table 1**: Distribution of disease duration by case category

| Case category | Distribution of disease duration (year) |
|---|---|
| Treated, HIV-negative | Uniform (0.2–2) |
| Not treated, HIV-negative | Uniform (1–4) |
| Treated, HIV-positive | Uniform (0.01–1) |
| Not treated, HIV-positive | Uniform (0.01–0.2) |

**Table 2**: Incidence estimation based on *U/T*

| | U (*n*) | T (*n*) | Prevalence ($10^{-3}$) | Duration (year) | Incidence ($10^{-3} y^{-1}$) |
|---|---|---|---|---|---|
| Cambodia 2002 | 260 | 42 | 12 (10-15) | 2.9 (1.9-4) | 4 (2.5-5.8) |
| Cambodia 2011 | 205 | 80 | 8.3 (7.1-9.8) | 1.2 (0.8-1.6) | 6.7 (4.5-9.3) |
| Myanmar 2009 | 300 | 79 | 6.1 (5-7.5) | 1.8 (1.1-1.6) | 3.3 (2-4.8) |
| Thailand 2012 | 136 | 60 | 2.5 (1.9-3.5) | 1.1 (0.5-1.6) | 2.3 (1-3.5) |

**Table 3**: Estimates of incidence derived from prevalence survey results, based on two estimation methods.

|  | Prevalence ($10^{-3}$) | Incidence - Method 1 ($10^{-3}y^{-1}$) | Incidence - Method 2 ($10^{-3}y^{-1}$) |
| --- | --- | --- | --- |
| Cambodia 2002 | 12 (10-15) | 4 (2.5-5.8) | 2.2 (1.5-2.9) |
| Cambodia 2011 | 8.3 (7.1-9.8) | 6.7 (4.5-9.3) | 3.8 (2.2-5.8) |
| Myanmar 2009 | 6.1 (5-7.5) | 3.3 (2-4.8) | 3.5 (2-5.1) |
| Thailand 2012 | 2.5 (1.9-3.5) | 2.3 (1-3.5) | 1.1 (0.7-1.6) |

**Table 4**: Distribution of CFRs by case category

|  | CFR | Sources |
| --- | --- | --- |
| Not on TB treatment $f_u$ | 0.43 (0.28-0.53) | 45,46 |
| On TB treatment $f_t$ | 0.03 (0-0.07) | 47 |

**Table 5**: Distribution of CFR in HIV-positive individuals

| ART | TB treatment | CFR | Sources |
|---|---|---|---|
| off | off | 0.78 (0.65-0.94) | [45] |
| off | on | 0.09 (0.03-0.15) | [47,48] |
| < 1 year | off | 0.62 (0.39-0.86) | Data from review + assumptions |
| < 1 year | on | 0.06 (0.01-0.13) | Data from review + assumptions |
| ≥ 1 year | off | 0.49 (0.31-0.70) | Assumptions |
| ≥ 1 year | on | 0.04 (0.00-0.10) | Assumptions |

# Figures

**Fig. 1** Main method to estimate TB incidence. In the first method, case notification data are combined with expert opinion about case detection gaps (under-reporting and under-diagnosis), and trends are estimated using either mortality data, repeat surveys of the annual risk of infection or exponential interpolation using estimates of case detection gaps for three years. For all high-income countries except the Netherlands and the United Kingdom, notifications are adjusted by a standard amount or measures of under-reporting from inventory studies, to account for case detection gaps.

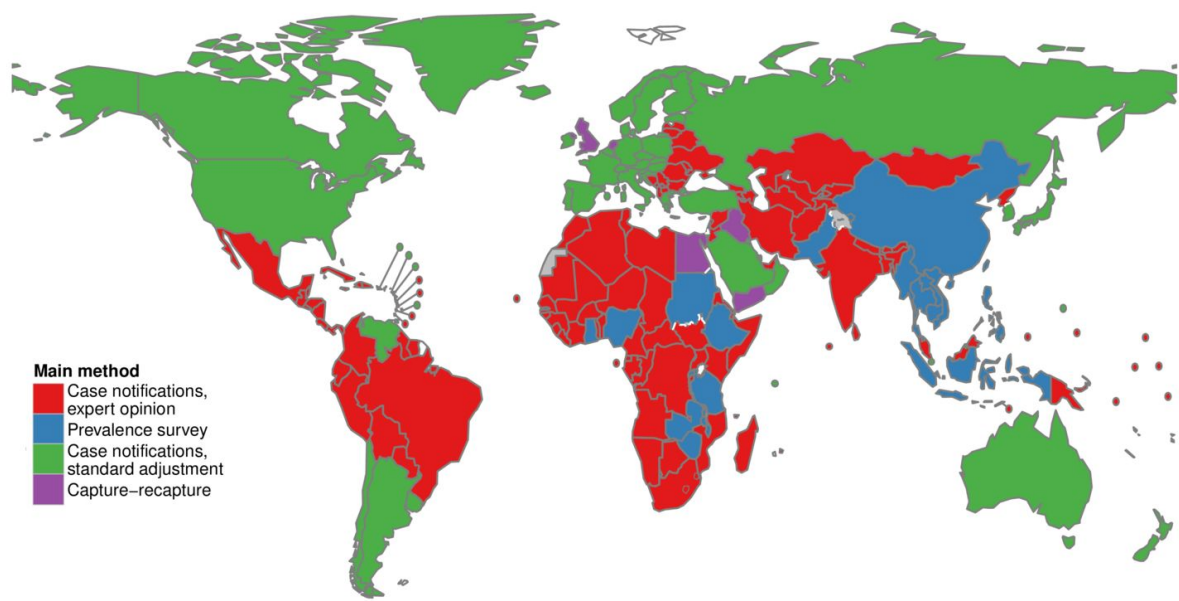

**Fig 2.** Estimates of TB incidence obtained indirectly (in blue) based on case notifications and expert opinion about detection and reporting gaps prior to a recent national prevalence survey, and derived from prevalence using survey results (in red), corresponding to the survey year (2012-2014). The length of segments indicate uncertainty ranges. Countries are presented in the order of the difference between estimates, grouped by continent. Estimates derived from prevalence are more robust. Post-survey estimates resulted in significant increases in estimated incidence in Nigeria and Indonesia, two countries with large population sizes resulting in a notable impact on global estimates. The often narrow uncertainty ranges of pre-survey estimates reflect a certain level of overconfidence of experts asked to estimate plausibility ranges for incidence. The post-survey estimate in Tanzania is highly uncertain, reflecting data management problems at the time of the survey as well as incomplete laboratory data.

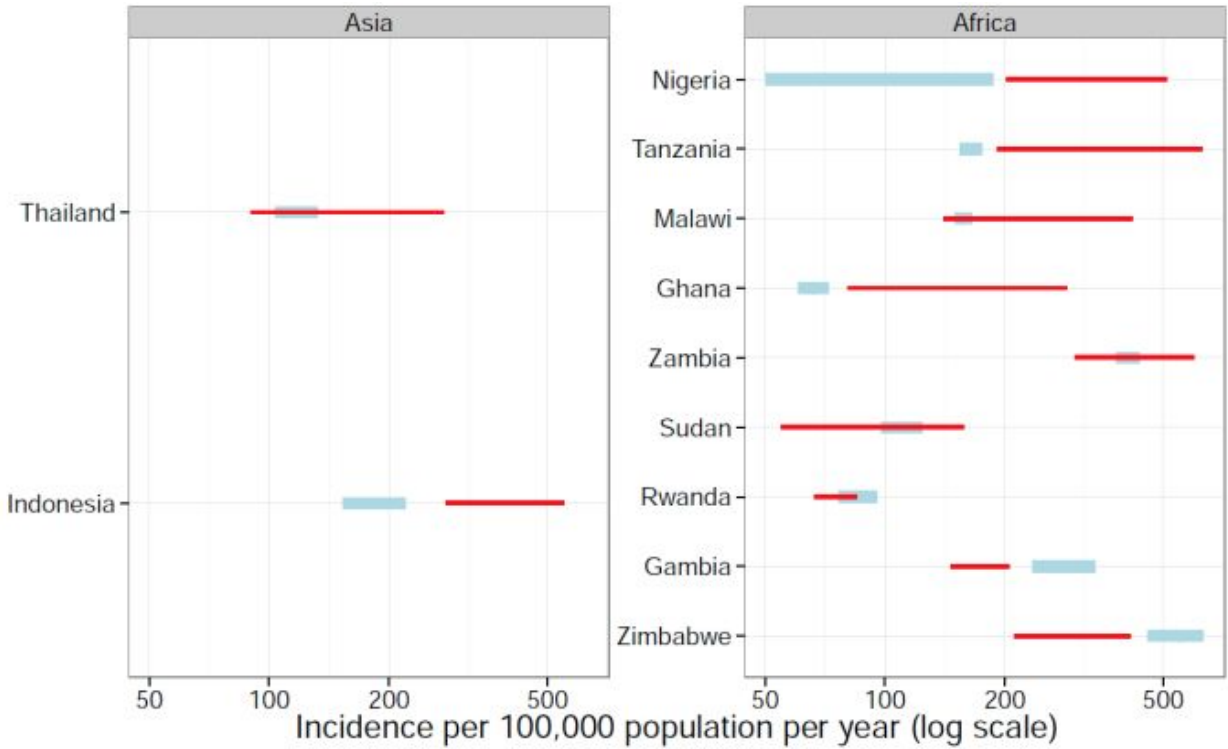

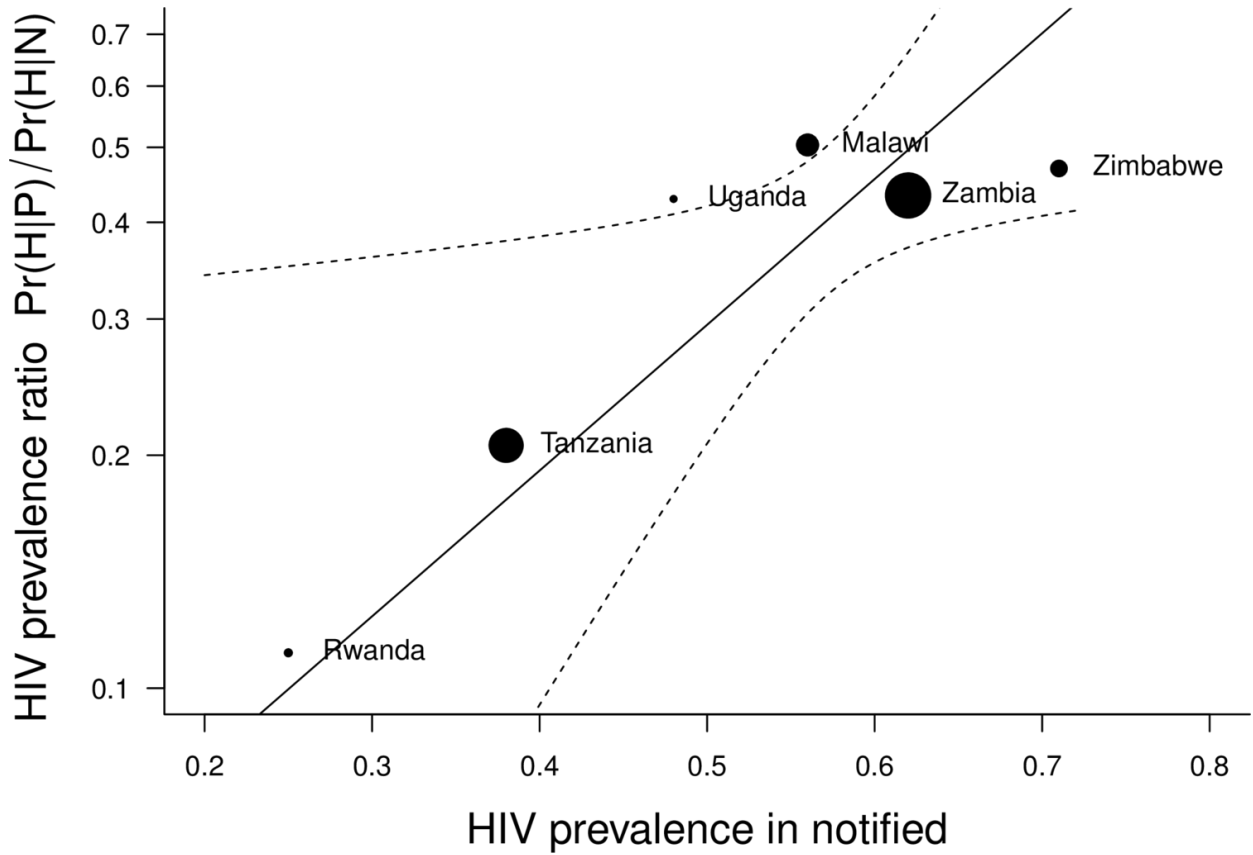

**Fig. 3** HIV prevalence rate ratio among prevalent TB cases over notified TB cases in 5 countries (points with area proportional to inverse variance) against HIV prevalence in notified cases. Predicted values of the prevalence rate ratio are shown with a solid line. The dashed lines represent 95% confidence bounds.

**Fig. 4** Reporting of new and relapse TB case notifications disaggregated by age, 2014

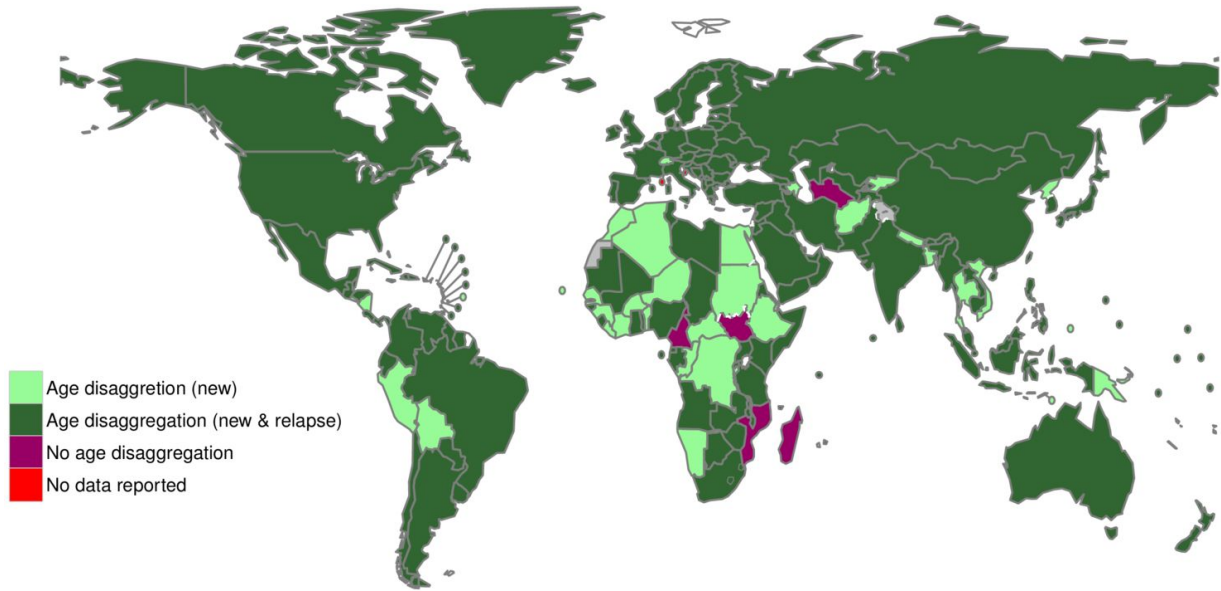

**Fig. 5** Countries for which prevalence is estimated from prevalence survey measurements (*n*=21). In the case of India, data from subnational surveys were pooled to provide a national estimate.

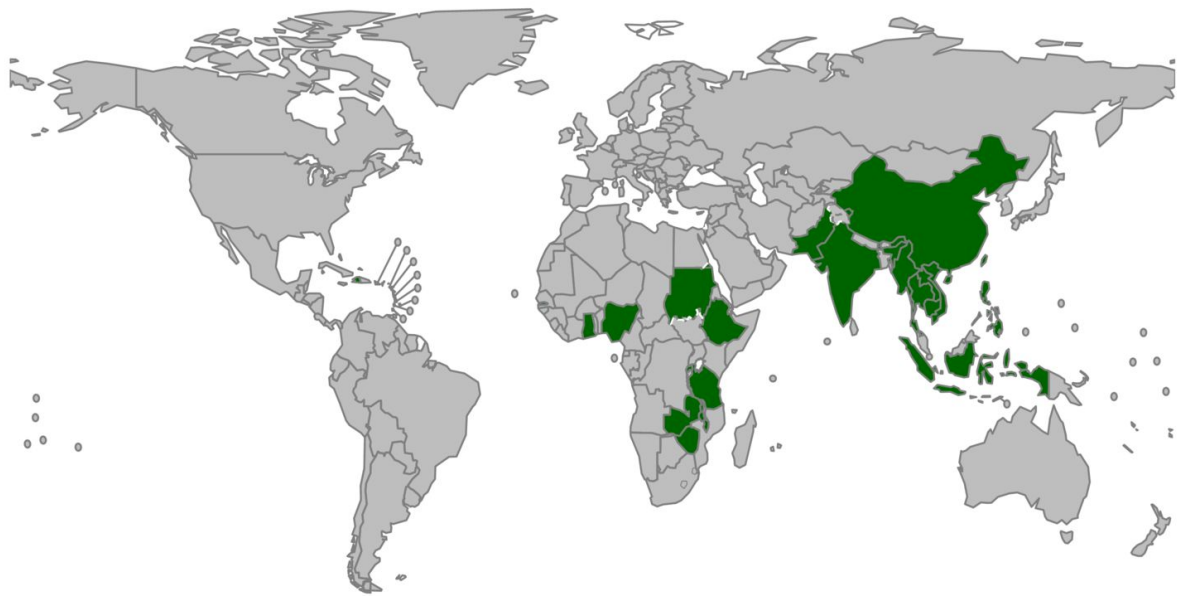

**Fig. 6** Countries (in green) for which TB mortality is estimated using measurements from vital registration systems (*n*=127) and/or mortality surveys (*n*=2)

**Fig. 7** Comparison of VR mortality (HIV-negative), horizontal axis (log scale) and mortality predicted as the product of incidence and CFR, vertical axis (log scale). Horizontal and vertical segments indicate uncertainty intervals. The dashed red line shows equality. The blue line and associated grey banner show the least-squared best fit to the data, with a slope not constrained to one.

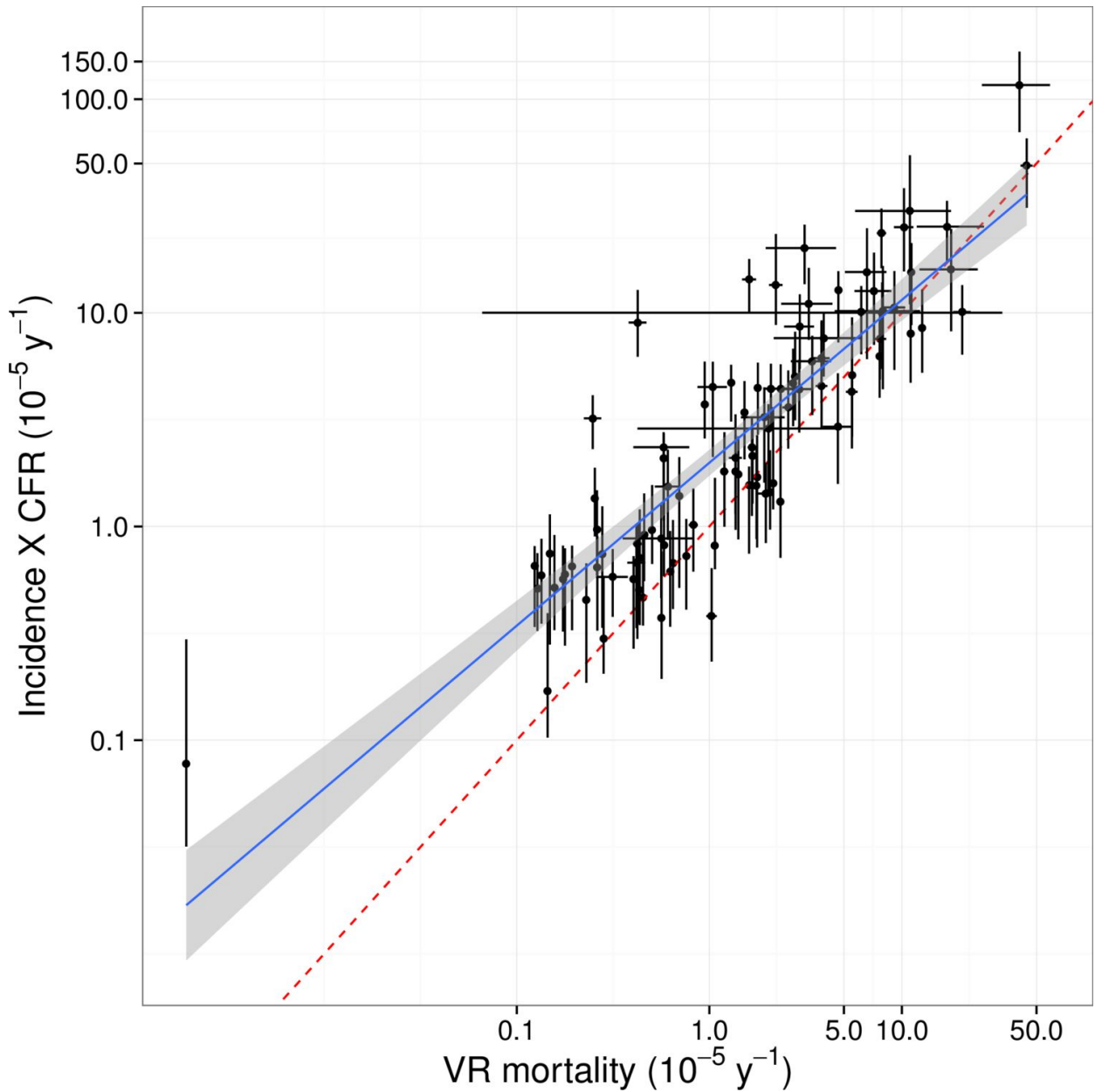